\documentstyle[floats,multicol,aps,epsf,prl]{revtex}

\begin{document}
\draft
\twocolumn[\hsize\textwidth\columnwidth\hsize\csname
@twocolumnfalse\endcsname
\def\btt#1{{\tt$\backslash$#1}}
\title{ Edge and Bulk Transport in the Mixed State of a Type-II Superconductor}
\author{Z. L. Xiao and E. Y. Andrei}
\address{Department of Physics and Astronomy, Rutgers University,
Piscataway, New Jersey 08855}
\author{Y. Paltiel and E. Zeldov}
\address{Department of Condensed Matter Physics, The Weizmann Institute of Science, Rehovot 76100, Israel}
\author{P. Shuk and M. Greenblatt} 
\address{Department of Chemistry, Rutgers University, Piscataway, New
Jersey 08854}
\maketitle
\begin{abstract}

{By comparing the voltage-current ($V-I$) curves obtained before and after 
cutting a sample of 2H-NbSe$_2$, we separate the bulk and edge contributions to 
the 
transport current at various dissipation levels and derive their respective $V-
I$ curves and critical currents. We find that the edge contribution is thermally 
activated across a current dependent surface barrier. By contrast the bulk $V-I$
curves are linear, as expected from the free flux flow model. The relative 
importance of bulk and edge contributions is found to depend on dissipation 
level and sample dimensions. We further show that the peak effect is a sharp 
bulk phenomenon and that it is broadened by the edge contribution.} 

\end{abstract}
]

{\bf I. INTRODUCTION}

One of the remarkable differences between a normal metal and a superconductor 
stems from the way they carry current. While in metallic samples the current 
density is usually homogeneous throughout the material, in superconductors in 
the Meissner state the current flows along the sample surface and edges so as to 
eliminate the self-induced magnetic field within the sample volume. In type-II 
superconductors in the mixed state, where the material is permeated by magnetic 
flux lines (vortices), a similar current enhancement along sample edges is the 
result of a surface barrier which inhibits the entry or exit of vortices 
\cite{1,2,3,4,5,6,7,8,9,10,11,12,13}. Several types of surface barriers have been identified 
including the Bean-Livingston  \cite{1} and geometrical \cite{2,3} barriers. The 
former, which is the primary source of edge currents in the experiments 
described here, is a result of the competition between the attraction of a 
vortex to its image and the 
repulsion arising from its interaction with shielding currents. Recent Hall 
probe measurements in Bi$_2$Sr$_2$CaCu$_2$O$_8$ \cite{4,5,6} and NbSe$_2$ \cite{7} 
crystals have shown  edge current enhancement due to 
the surface barrier. But thus far these edge currents were only qualitatively 
identified. 

The experiments described here allow for the first time to derive individual 
voltage-current ($V-I$) characteristics of the edge and bulk currents. Below the 
peak effect region (a peak in the critical current just below $T_c$) we find 
that vortex entry and exit at the edges is governed by thermal activation across 
a current dependent surface barrier, whereas vortex motion in the bulk is non-
activated. Our results 
show that the observed nonlinearities of the $V-I$ characteristics are due to 
the edge contribution while the bulk $V-I$ curves are linear. We further show 
that the peak in critical current is primarily a bulk effect which sharpens and 
becomes more pronounced  when 
the edge contribution is removed. These experiments demonstrate that boundaries 
have a profound effect on the $V-I$ characteristics as well as on the field and 
temperature dependence of the critical current.

{\bf II. EXPERIMENTAL DETAILS}

The experiments were carried out in the low temperature superconductor 2H-NbSe$_2$ where local self-field \cite{7} and magnetization \cite{13} measurements 
unveiled the 
presence of significant edge currents. Transport, neutron scattering and 
magnetization measurements in these samples \cite{14,15,16,17,18,19,20,21,22,23,24,25,26} revealed 
a number of unusual phenomena including memory and current
driven reorganization \cite{16,17,18,19,20,21,22,23,24}. In addition, the shape of the peak 
effect in this material was found to change significantly with measurement speed 
\cite{20,24} or 
contact configuration \cite{25}. Several of these results were shown to be a 
consequence of vortices traversing a surface barrier as they enter the sample 
\cite{24,25}.  In the experiments described here the surface barrier was 
determined 
after separating the edge and bulk contributions to the current. The separation 
procedure involves cutting a sample to reduce its width and comparing the $V-I$
characteristics before and after cutting. A schematic illustration of the 
cutting is given in Fig. 1(a).

The data were acquired on two undoped single crystals of 2H-NbSe$_2$ with 
initial sizes of  $8(L)$ $\times$ $1.72(w)$ $\times$ $0.020(d)$ mm$^3$ (sample 
A) and $6.3(L)$ $\times$ $1.40(w)$ $\times$ $0.060(d)$ mm$^3$ (sample B) and 
with zero field critical temperatures and width of $T_c=$ 7.18 K, 
$\Delta T_c=$ 95 mK and 7.21 K, 92 mK, respectively. Our measurements employed a 
standard four probe technique with low resistance Ag$_{0.1}$In$_{0.9}$ solder 
contacts. The distance between the voltage contacts was 2.5 mm and 1.5 mm in 
samples A and B, respectively. The critical current, $I_c$, is defined as the 
current at which the 
voltage reaches 1 $\mu$V. The magnetic field was kept along the $c$ axis of the 
sample 
and the dc current was applied in the $a-b$ plane. The vortex lattices are 
prepared by applying the magnetic field after cooling the sample through  $T_c$ 
(zero-field  
\begin{figure}[btp]
\epsfxsize=3.8 in
\epsfbox{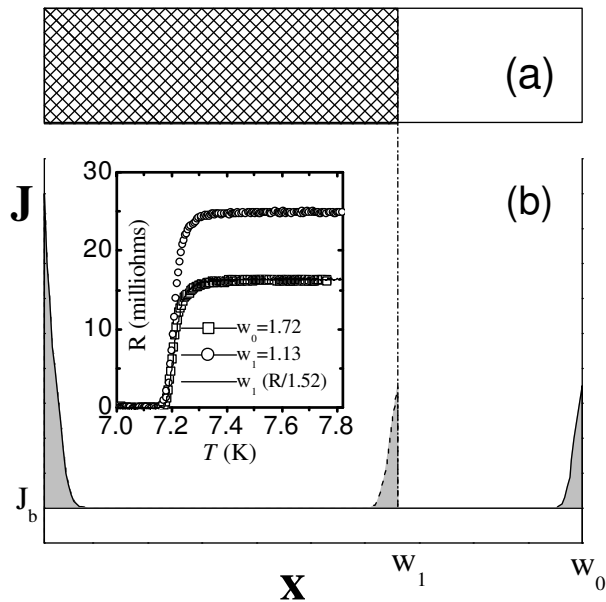} 
\protect\caption{Schematic illustrations of sample cutting (a) and the 
corresponding current distributions in the vortex state (b).The shadowed area 
represents the edge current $I_s$ (see text).  The inset in (b) 
shows the resistance versus temperature curves and the determination of the 
ratio of the sample  width before ($w_0$) and after ($w_1$) cutting. }
\label{fig:fig_pl1}
\end{figure}
cooling). Sample cut was carried out manually with a sharp razor blade. The 
width reduction factor $\alpha_{0n}=w_0/w_n$  ($w_0$, $w_n$ are the width before 
and after $ n^{th}$ cutting) was determined by direct inspection under a 
microscope and confirmed by measuring the ratio of normal state resistances 
before and after cutting. Sample A was cut twice with $\alpha_{01}= $ 1.52 
[shown in the inset of Fig. 1(b)] and $\alpha_{02}=$ 3.65 for the first and 
second cuts respectively, and sample B was cut once with $\alpha_{01}=$ 1.35. By 
using this procedure rather than samples with different widths one can be 
certain that in comparing the $V-I$ curves before and after cutting all the 
parameters (excepting the newly cut edge) are the same.

{\bf III. RESULTS AND DISCUSSION}

In Fig. 2 we show the effect of sample cutting by comparing the $V-I$ curves and 
critical currents before and after cutting. In Fig. 2(a) the results are shown 
for both the normal ($T=$ 7.6 K $>$ $T_c$) and superconducting ($T=$ 4.25 K $<  
T_c$) states at $H=$ 1 T. 
A comparison of the measurements at the same average current density  ($J=I/dw$) 
is obtained by plotting the voltage against the scaled current $\alpha_{02} I$  
(solid line). At 7.6 K the scaled curve exactly overlaps the response in the 
uncut 
sample, clearly showing that the current density in the normal state scales with 
the inverse of the sample width. In other words the current distribution in the 
normal state is uniform.  Using the same procedure for the data in the vortex 
state, we find that the scaled $V-I$ curve of the cut sample is shifted to the 
right of the initial curve,
\begin{figure}[btp]
\epsfxsize=3.5in
\epsfbox{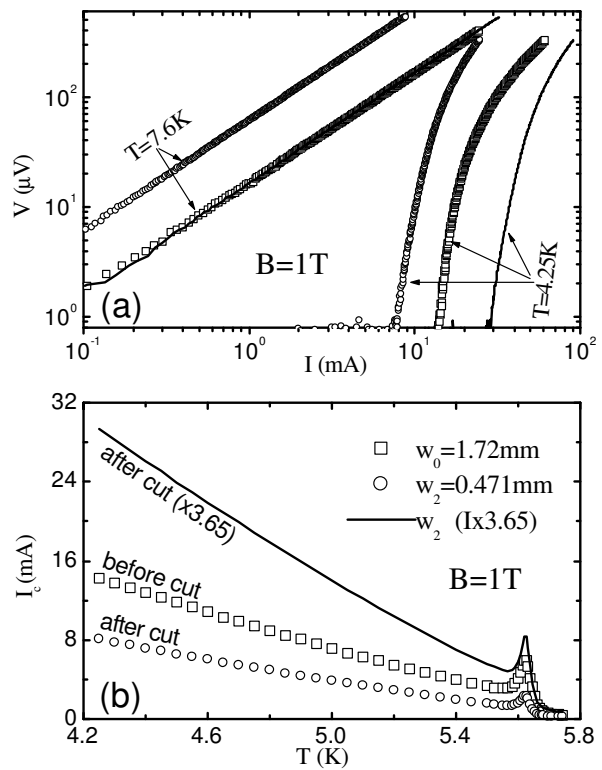} 
\protect\caption{$V-I$ curves at temperatures below (4.25 K) and above (7.60 K) 
the critical 
temperature $T_c$ (a) and temperature dependence of critical current (b) for 
sample A before and after 
sample cutting. The open symbols and solid lines represent the as-measured and  
the scaled data.} \label{fig:fig_pl1}
\end{figure}
indicating that for the same voltage response the 
average current density in the cut sample is much higher. The same tendency is 
found for the scaled critical currents, which are significantly larger after 
cutting, Fig. 2(b). As shown below these results are consistent with an enhanced 
current at the sample edges due to the surface barrier.  In fact both theories 
\cite{1,10,11} and experiments\cite{4,5,6,7,12,13} favor an excess current carrying 
capacity at the sample edges. The results are analyzed by separating the total 
current, $I=I_s+I_b$, into a 
uniformly distributed bulk contribution, $I_b=J_bwd$, with current density 
$J_b$ and a 
nonuniform contribution, $I_s$. The latter, assumed to remain unchanged after 
cutting, represents the edge current due to the surface barrier\cite{1,2,3}. If 
edge contamination \cite{24} is present the resulting enhanced edge current would 
also be included in $I_s$. This model is applicable when the sample width after 
cutting is 
much larger than the characteristic extent of edge currents. In our experiments 
this condition is well satisfied below the peak effect regime where edge 
currents are predominantly due to the surface barrier  (edge contamination is 
negligible 
\cite{24})  and are thus confined to within a narrow strip $ \le \lambda \sim 100 
nm \ll  w $ \cite{10,11}. It follows that the vortex velocity
\begin{figure}[btp]
\epsfxsize=3.5in
\epsfbox{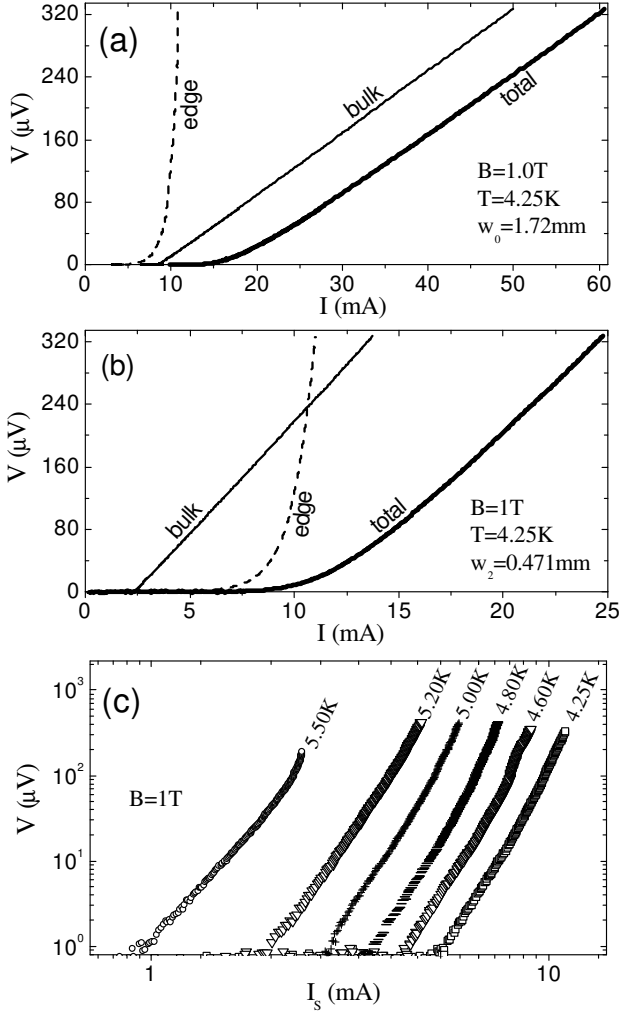} 
\protect\caption{$V-I$ curves for total  (thick line), edge (dashed) and bulk 
(thin line) current before (a) and after (b) cutting the sample. (c) edge $V-
I_s$ curves in 1 T and at various temperatures  }
\label{fig:fig_pl1}
\end{figure}
and hence the voltage response at a given field and temperature is uniquely 
determined by the bulk current density $J_b$ and by $I_s$. 
In Fig. 1(b) we plot a schematic current distribution in the presence of a 
surface 
barrier \cite{1,10,11} or/and edge contamination illustrating the effect of cutting 
for a constant $J_b$. In this model the voltage response to a driving current 
$I=I_b+I_s$ in 
the initial sample will be the same as the response to a current 
$I_\alpha=I_b/\alpha+I_s$ in 
the cut sample, leading to a straightforward procedure for analyzing the data 
and for separating bulk from surface currents. Thus, after obtaining the values 
of $I$ and $I_\alpha$ at a given voltage response by measuring the $V-I$ curves 
before and after cutting the sample, the bulk and surface contributions to the 
current are given by: $I_b =\alpha (I-I_\alpha)/(\alpha-1)$ and $I_s=(\alpha 
I_\alpha-I)/(\alpha-1)$. Repeating the same procedure 
for various voltage levels, separate $V-I$ characteristics can be obtained for 
the bulk ($V-I_b$) and the edge ($V-I_s$). Results for the $V-I$ characteristics 
of sample A at 4.25 K and 1 T are 
\begin{figure}[btp]

\epsfxsize=3.5in
\epsfbox{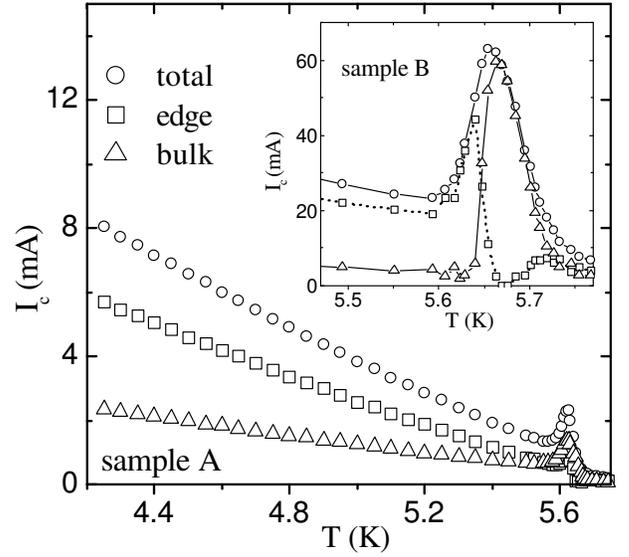} 
\protect\caption{Temperature dependence of total, edge and bulk critical 
currents at 1 T.  Main panel:  results for sample A ($w_2=$ 0.471 mm); inset: 
expanded view of the peak region for sample B ($w_0=$ 1.40 mm).  }
\label{fig:fig_pl1}
\end{figure}
shown in Fig. 3 (a) and (b) for the initial ($w_0=$ 1.72 mm) and cut sample 
($w_2=$ 0.471 mm) respectively. We note that the relative contribution 
of the edge is larger in the narrower sample Fig. 3(b) than in the wider one 
Fig. 3(a), in accord with the assumption that $I_s$ is unchanged, whereas $I_b$ 
is 
proportional to the sample width. In addition, the relative contribution of the 
edge to the total transport current diminishes with increasing dissipation level 
indicating that the voltage grows much faster with increasing $ I_s$ than it 
does 
with $I_b$. We further note that the measured $V-I$ curves in both the initial 
and cut samples are nonlinear at modest dissipation levels. Previous 
interpretations of the nonlinearities in the $V-I$ curves in 2H-NbSe$_2$ were 
usually  based on the assumption of a bulk phenomenon.  However from Fig. 3(a) 
and (b), our data show  that the nonlinearity is due to the edge contribution 
whereas the bulk $V-I$ curves are remarkably linear, consistent with early 
models of vortex motion in low temperature superconductors\cite{27,28}. 

Theoretical studies of vortex entry and exit across sample edges \cite{10,11} have 
shown 
that a surface barrier $U(I_s)$, gives rise to thermally activated $V-I_s$
characteristics: $V=V_0exp[-U(I_s)/k_BT]$. The current dependence of the surface 
barrier is determined by the mechanism of vortex penetration. For example, the 
surface barrier in a slab geometry was found to have a power law current 
dependence, $U \sim I_s^{-1/2}$, in the 3D case while in the 2D case it is 
logarithmic $U \sim U_oln(I_o/I_s)^{1/2}$  with $U_o$ a characteristic energy 
scale \cite{10,11}.  In Fig. 3(c) we plot the 
edge $V-I_s$ characteristics at 1 T for several temperatures. The data are best 
fitted with a logarithmic current dependence of the surface barrier $U(I_s)$, as 
seen from the straight lines obtained on a log-log scale. The temperature 
dependence of the slope is consistent with thermal activation with $U_o \sim$ 70 
K for $T<$ 5 K and decreasing in value for $T>$ 5 K.

The temperature dependence of the critical current for samples A and B is shown 
in Fig. 4. The as-measured critical currents are shown together with the bulk 
($I_{cb}$) and edge ($I_{cs}$) contributions obtained by the procedure described 
above.  
The two samples were grown in the same batch and are expected to be similar in 
quality, but they differ in thickness, with sample B three times thicker than 
sample A. Comparing the bulk critical current densities ($J_{cb}=I_{cb}/wd)$ in 
the two 
samples we find  that below the peak effect they are practically identical, 
despite the fact that the average critical current density, $J_c=I_c/wd$, 
calculated 
in the usual way as the total critical current divided by the sample cross 
section is more than 3  times larger in sample B than in A. These data show 
that the two samples are identical in their bulk pinning properties and that the 
distribution of pinning centers in the bulk is homogeneous. The significant 
difference between the bulk and average critical current densities must 
therefore be due to the surface barrier. Comparing the surface critical sheet 
current densities, defined as $J_{cs} =I_{cs}/d$,  we find a 
strong enhancement of the edge critical current 
density in sample B compared to sample A, which implies that 
the surface barrier in the thicker sample is larger. 
More work is needed to understand the dependence of the surface barrier on  
sample thickness.

We next consider the peak effect region \cite{29}, where both samples exhibit a 
well 
defined enhancement in the total critical current by factors of 2.0 and 2.7 for 
sample A and B respectively. The shape of the peak was previously found to 
depend on measurement speed and contact geometry becoming much sharper and more 
pronounced when the contribution from vortices entering through the edges was 
reduced or eliminated \cite{24,25}.  A similar result is obtained here by 
separating the edge and bulk contributions to the critical current.  As shown in 
the inset of Fig. 4 the {\it bulk}  peak effect consists of a 20 fold 
enhancement in 
$I_{cb}$ [$I_{cb}$($T=$ 5.63 K)$=$ 2.9 mA; $I_{cb}$($T=$ 5.66 K)$=$ 59.6 mA] 
over a temperature range narrower 
than the width of the zero field superconducting transition. This remarkably 
sharp peak is similar to that obtained when vortex crossing of edges is 
eliminated by using a Corbino geometry \cite{25}. In the lower part of the peak $I_{cs}$ increases rapidly 
with 
temperature compared to its monotonic decrease below the peak. This enhancement 
is consistent with the edge-contamination mechanism suggested by Paltiel et al. 
\cite{24} since the higher critical current of the  disordered phase, injected at 
the sample edge in the lower part of the peak effect, leads to an additional 
contribution to  $I_{cs}$ over and above that due to the surface barrier. 
These results  are also consistent with the significant sharpening of the peak 
effect observed in high frequency ac measurements \cite{18,19,24} where edge 
contamination is practically eliminated.

In the above analysis we assumed that the edge contribution is unchanged by 
cutting the sample. As a check we cut the sample a second time and compared the 
bulk $V-I$ characteristics to those before cutting. Below the peak region we 
found that the $V-J_b$ curves were unchanged within better than 10\%.

{\bf IV. CONCLUSIONS}

In conclusion, by comparing the $V-I$ curves of samples before and after they 
are 
cut, we obtained the bulk and edge contributions to the transport current at 
various dissipation levels. This led to the first derivation of separate $V-I$ 
curves for the bulk and the edge. Below the peak region the nonlinearity of the measured 
$V-I$ curves in 2H-NbSe$_2$ is due to the edge contribution and 
the edge current is governed by thermally activated vortex crossing 
through a current dependent surface barrier. By contrast the bulk $V-I$ 
characteristics are linear confirming the free flux flow model for vortex motion 
in the bulk. In the peak effect region the temperature dependence of the bulk 
critical current exhibits a very sharp peak.  The edge contribution starts increasing 
before the bulk current does, leading to a smeared out peak in the total 
critical current.

{\bf ACKNOWLEDMENTS}

Work supported by NSF-DMR 97-05389 and by DOE DE-FG02-99ER45742. EZ acknowledges 
support by the German-Israeli Foundation G.I.F.


\begin{references}
 \bibitem{1} C. P. Bean and J. D. Livingston, Phys. Rev. Lett. {\bf 12}, 14 (1964). 
\bibitem{2} E. Zeldov, A. I. Larkin, V. B. Geshkenbein, M. Konczykowski, D. Majer, B. Khaykovich, V. M. Vinokur, and  H. Shtrikman,  Phys. Rev. Lett. {\bf 73}, 1428 (1994).
\bibitem{3} M. Benkraouda and J. R. Clem, Phys. Rev. B {\bf 58}, 15103 (1998). 
\bibitem{4}  Dan T. Fuchs, Eli Zeldov, Michael Rappaport, Tsuyoshi Tamegai, Shuuichi Ooi, and Hadas Shtrikman, Nature {\bf 391}, 373 (1998). 
\bibitem{5} D. T. Fuchs, E. Zeldov, T. Tamegai, S. Ooi, M. Rappaport, and H. Shtrikman, Phys. Rev. Lett. {\bf 80}, 4971 (1998).
\bibitem{6} D. T. Fuchs, R. A. Doyle, E. Zeldov, S. F. W. R. Rycroft, T. Tamegai, S. Ooi, M. L. Rappaport, and Y. Myasoedov,  Phys. Rev. Lett.  {\bf 81}, 3944 (1998).
\bibitem{7} Y. Paltiel, D. T. Fuchs, E. Zeldov, Y. N. Myasoedov, H. Shtrikman, M. L. Rappaport, and  E. Y. Andrei, Phys. Rev. B {\bf 58}, R14763 (1998).
\bibitem{8} M. Konczykowski, L. I. Burlachkov, Y. Yeshurun, and F. Holtzberg, Phys. Rev. B {\bf 43}, 13707 (1991).
\bibitem{9} N. Chikumoto, M. Konczykowski, N. Motohira, and A. P. Malozemoff, Phys. Rev. Lett. {\bf 69}, 1260 (1992).
\bibitem{10} L. Burlachkov, V. B. Geshkenbein, and A. E. Koshelev, Phys. Rev. B {\bf 50}, 16770 (1994).
\bibitem{11} L. Burlachkov, A. E. Koshelev, and V. M. Vinokur, Phys. Rev. B {\bf 54}, 6750 (1996).
\bibitem{12} S. F. W. R. Rycroft, R. A. Doyle, D. T. Fuchs, E. Zeldov, R. J. Drost, P. H. Kes, T. Tamega, S. Ooi, and  D. T. Foord, Phys. Rev. B {\bf 60}, R757 (1999).
\bibitem{13} P. K. Mishra, G. Ravikumar, T. V. C. Rao, V. C. Sahni, S. S. Banerjee, S. Ramakrishnan, A. K. Grover, and M. J. Higgins,  Physica C {\bf 340}, 65 (2000). 
\bibitem{14}  M. J. Higgins and S. Bhattacharya, Physica C {\bf 257}, 232 (1996). 
\bibitem{15} X. S. Ling,  J. E. Berger, and  D. E. Prober, Phys. Rev. B {\bf 57}, R3249 (1998).

\bibitem{16}   W. Henderson, E. Y. Andrei, M. J. Higgins,  and S. Bhattacharya, 
Phys. Rev. Lett. {\bf 77}, 2077 (1996).
\bibitem{17}  Z. L. Xiao, E. Y. Andrei, P. Shuk,  and M. Greenblatt, Phys. Rev. Lett.  {\bf 85}, 3265 (2000).
\bibitem{18} W. Henderson, E. Y. Andrei,  and M. Higgins, Phys. Rev. Lett. {\bf 81}, 2352 (1998).
\bibitem{19} E. Y. Andrei, Z. L. Xiao, W. Henderson, M. J. Higgins, P. Shuk, and M. Greenblatt, J. Phys. IV, {\bf Pr10} 5 (1999). 
\bibitem{20} Z. L. Xiao, E. Y. Andrei, and M. J. Higgins, Phys. Rev. Lett. {\bf 83}, 1664 (1999).
\bibitem{21} U. Yaron, P. L. Gammel, D. A. Huse, R. N. Kleinman, C. S. Oglesby, E. Bucher, B. Batlogg, D. J. Bishop, K. Mortensen, and K. N. Clausen, Nature {\bf 376}, 753 (1995).
\bibitem{22} F. Pardo, F. de la Cruz, P. L. Gammel, E. Bucher, and  D. J. Bishop, Nature {\bf 396}, 348 (1998)
\bibitem{23} Z. L. Xiao, E. Y. Andrei, P. Shuk, and M. Greenblatt, Phys. Rev. Lett. {\bf 86}, 2431 (2001).
\bibitem{24} Y. Paltiel, E. Zeldov, Y. N. Myasoedov, H. Shtrikman, S. Bhattacharya, M. J. Higgins, Z. L. Xiao, E. Y. Andrei, P. L. Gammel, and D. J. Bishop, Nature {\bf 403}, 398 (2000).
\bibitem{25} Y. Paltiel, E. Zeldov, Y. Myasoedov, M. L. Rappaport, G. Jung, S. Bhattacharya, M. J. Higgins, Z. L. Xiao, E. Y. Andrei, P. L. Gammel, and D. J. Bishop, Phys. Rev. Lett. {\bf 85}, 3712 (2000). 
\bibitem{26} S. S. Banerjee, N. G. Patil, S. Saha, S. Ramakrishnan, A. K. Grover, S. Bhattacharya, G. Ravikumar, P. K. Mishra, T. V. Chandrasekhar Rao,  V. C. Sahni,  M. J. Higgins, E. Yamamoto, Y. Haga,  M. Hedo, and Y. Inada, Phys. Rev. B {\bf 58}, 995 (1998).
\bibitem{27} Y. B. Kim, C. F. Hempstead, and A. R. Strnad, Phys. Rev. {\bf 139}, 
A1163 (1965)
\bibitem{28} J. Bardeen and M. J.  Stephen, Phys. Rev. {\bf 140}, A1197 (1965). 
\bibitem{29} In the peak effect region the current separation is sensitive to 
the position of the peak. A small shift in the peak, due for example to a 
misalignment between the $c$ axis of the sample and the field, can lead to 
different relative magnitudes of edge and bulk contributions. In spite of these 
variations, we found that the peak in the bulk contribution was always 
significantly sharper than that in the total current.

\end{references}
\end{document}